\documentclass[aps,prl,twocolumn,showpacs,nobibnotes]{revtex4-1}
\usepackage{graphics,graphicx,amsfonts,amsmath,amsbsy,amssymb,color}
\usepackage[singlelinecheck=false]{subfig}
\usepackage{verbatim}  
\usepackage{psfrag}  
\usepackage{tabularx}
\usepackage{xmpmulti}

\def \beq {\begin{eqnarray}}
\def \eeq {\end{eqnarray}}

\newcommand{\Hamil}{\hat{H}}

\newcommand{\reffig}[1]{{Fig.~\ref{#1}}}

\newcommand{\half}{\frac{1}{2}}

\begin{document}

\title{Many-body quantum chemistry for the electron gas: convergent perturbative theories}

\author{James~J.~Shepherd}
\email{js615@cam.ac.uk}
\affiliation{University of Cambridge, The University Chemical Laboratory, \\Lensfield Road, Cambridge, CB2 1EW, United Kingdom}
\author{Andreas~Gr\"{u}neis}
\email{andreas.grueneis@univie.ac.at}
\affiliation{University of Vienna, Faculty of Physics and Center for Computational Materials Science, Sensengasse 8/12, A-1090 Vienna, Austria}

\begin{abstract}
We investigate the accuracy of a number of wavefunction based methods at the heart of quantum chemistry
for metallic systems.
Using Hartree--Fock as a reference, perturbative (M\o ller-Plesset, MP) and coupled cluster (CC) theories are
used to study the uniform electron gas model.
Our findings suggest that non-perturbative coupled cluster
theories are acceptable for modelling electronic interactions in metals whilst perturbative coupled cluster theories
are not.
Using screened interactions, we propose a simple modification 
to the widely-used coupled-cluster singles and doubles plus
perturbative triples method (CCSD(T)) that lifts the divergent behaviour and
is shown to give very accurate correlation energies for the homogeneous electron gas. 
\end{abstract}
\date{\today}
\maketitle

\emph{Introduction.} -- 
The accurate solution of the many-electron Schr\"odinger equation for real solid state systems
is a major challenge for the field of theoretical solid state physics and holds the key to understanding
materials with some of the most intriguing and commercially valuable properties, such as high T$_{\rm c}$
cuprates and transition metal oxides.
Moreover, it provides access to properties of materials under conditions
that are not accessible by experiment.
During the last years a number of quantum chemical
methods such as second-order M\o ller-Plesset (MP2) and coupled cluster singles and doubles (CCSD) theory
have made significant progress towards finding accurate solutions for real materials~
\cite{Booth2013,Ben2012,Paulus2012,Schutz2007,Pisani2007,Scuseria2001,Schutz2011,Kresse2009,Stoll2012,Manby2009,Gillan2010,Schwerdtfeger2010}.
Thanks to recent methodological advances and the increase in available computer power, these methods have become
tractable for simple materials despite their computational cost.
It has been shown that coupled cluster theory, one of the most accurate and widely-used quantum chemical methods
translates its high accuracy and systematic improvability
seamlessly from molecular systems to semiconductors and insulators, including even more strongly correlated
systems such as NiO~\cite{Booth2013}.

The development of these and other highly accurate as well as predictive ab-initio methods is partly also motivated
by their potential use for studying metal-insulator transitions in transition metals oxides,
where currently available density functionals fail~\cite{Marsman2008}.
However, so far very little is known about the accuracy of quantum chemical wavefunction
based methods for metallic systems.
An open question of growing importance surrounding this field is to directly address 
which methods are appropriate and which are not for the study of metallic
systems. Approximations and divergences need to be understood so that needless effort is
not expended investigating methods which will ultimately fail.
Although it would in principle be possible to pursue this question with analytical theory, 
the plurality of diagrams and the lack of closed solutions makes this attempt intractable.
To this end we aim to provide here a simple, novel and robust methodology to test for the
numerical convergence of approximate methods in metals using the \emph{finite basis set}
simulation-cell electron gas~\cite{Shepherd2012a,Shepherd2012b,Shepherd2012c}.
Furthermore we propose modifications to the employed quantum chemical methods
that account for screening effects and correct deficiencies of
perturbative M\o ller-Plesset and coupled cluster theories in metals.

\emph{Theory.} -- 
In this work we will employ various quantum chemical methods that use Hartree--Fock as a reference
and treat electronic correlation by expanding the many-electron wavefunction in a multideterminantal
basis.
The electronic correlation energy in the MP2 and CCSD theory
is non-variational and can for the systems studied in this work be calculated by
\begin{equation}
E^{\rm c}=\frac{1}{4}\sum_{i,j}\sum_{a,b} t_{ij}^{ab} \langle \Psi_{\rm HF} | H | \Psi_{ij}^{ab} \rangle 
\label{eq:energy}
\end{equation}
The indices $i$,$j$,$k$,$l$ and $a$,$b$,$c$,$d$
will be used throughout this work to refer to occupied and unoccupied HF spin orbitals, respectively.
In the above expression $\Psi_{ij}^{ab}$ are HF Slater determinants where the occupied orbitals $i$ and $j$ have been
replaced with virtual orbitals $a$ and $b$. 
The $t_{ij}^{ab}$'s refer to the coefficients of the doubly excited  Slater determinants and their
definition for the various wavefunction based methods will be given later.
We note that singly excited Slater determinants have no contributions to the wavefunctions and correlation energies
of the systems studied in the present paper.

In this work we seek to investigate the accuracy of various wavefunction based methods for an archetypal
fully three dimensional metallic system.
The homogeneous electron gas (HEG, uniform electron gas or jellium model)
is well-regarded to be the simplest model for a metallic system, consisting of $N$
electrons in a box of length $L$ with a two-electron Ewald interaction
$\hat{v}_{\alpha\beta}$\cite{Ewald1921,Fraser1996},
\begin{equation}
\Hamil=\sum_\alpha -\half \nabla_\alpha^2 + \sum_{\alpha\neq \beta} \half \hat{v}_{\alpha\beta} + \text{const.}
\label{sim_cell_H}
\end{equation}
In the thermodynamic limit (TDL), found as the particle number tends to infinity
($N\rightarrow\infty$) with the density held constant, it is possible to solve
the above Hamiltonian in the Hartree--Fock approximation with plane waves.
This yields an analytic expression for the dispersion relation,
producing a band structure with a zero in the density of states
at the Fermi energy~\cite{Martin2004}; this complicates analytical derivations.

In setting out to find the behaviour of approximate theories to obtain the correlation energy
(i.e. the total energy with Hartree--Fock energy as a starting-point), it is typical to start
with a finite simulation-cell model of $N$ electrons, and carefully approach the thermodynamic
limit by extrapolation~\cite{Fraser1996,Drummond2008}.
However, in quantum chemical techniques,
we must also make do with a finite one-particle basis set.
The difficulty of investigating the properties of these approximate theories in the
thermodynamic limit is hampered by this requirement of a finite basis set, in this
case of $M$ plane waves spinorbitals defined by a kinetic energy cutoff
$\frac{1}{2} k_c^2$. In principle, the complete basis set limit $k_c\rightarrow\infty$
and thermodynamic limit $N\rightarrow\infty$ must be found, which is prohibitively costly
given the scaling of even approximate quantum chemical theories.

\begin{figure}[t]
\includegraphics[width=0.45\textwidth,clip=true]{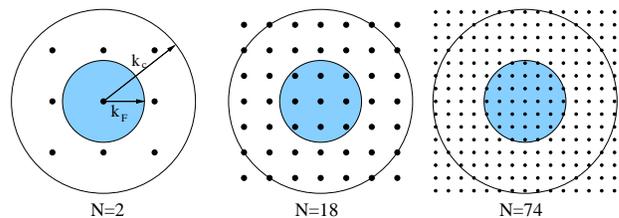}
\caption{(Colour online)
Schematic illustration of two dimensional reciprocal lattice for different electron numbers $N$ ($N$=2, 18 and 74) and
a fixed density.
$k_F$ and $k_c$ denote the length of the Fermi- and basis set cutoff wave vector, respectively.
The length of these wave vectors is constant for a fixed density.
}
\label{fig:1a}
\end{figure}

The most obvious way to make progress towards these limits is to take the
$k_c\rightarrow\infty$ limit, to solve the $N$-electron Hamiltonian at the
complete basis set limit, and then the $N\rightarrow\infty$ limit can be found latterly.
However, in this study we propose to take the $N\rightarrow\infty$ limit first for a finite
$k_c=\gamma k_{\rm F}$\footnote{Taking a \emph{finite-basis} approach to the homogeneous electron gas has
only recently been introduced to the literature~\cite{Shepherd2012a,Shepherd2012b,Shepherd2012c}}.
Figure~\ref{fig:1a} illustrates this approach schematically for a two dimensional reciprocal lattice.
As the $N\rightarrow\infty$ limit is taken, the band gap closes
because the grid spacing in the region around the Fermi surface becomes smaller, and the
zero-momentum excitations that cause the divergences in for example MP2
theory are increasingly well-represented in a size-extensive fashion.

\emph{Results.} -- We first outline how to show the well-known divergence in the MP2 energy using finite-$M$~\cite{Gruneis2010},
finite-$N$ calculations and then generalise this approach to demonstrate limiting behaviours
in other theories. 
The amplitudes in MP2 theory are given by
\begin{equation}
t_{ij}^{ab}
=\frac{\langle \Psi_{ij}^{ab} | \hat{v} | \Psi^{\rm HF} \rangle}{\epsilon_i+\epsilon_j-\epsilon_a-\epsilon_b},
\label{eq:mp2amplitudes}
\end{equation}
where $\epsilon$ are the Hartree--Fock eigenvalues of the spin orbitals, and are employed for the evaluation of the MP2 correlation energy according to Eq.~(\ref{eq:energy}).

\begin{figure*}[t]
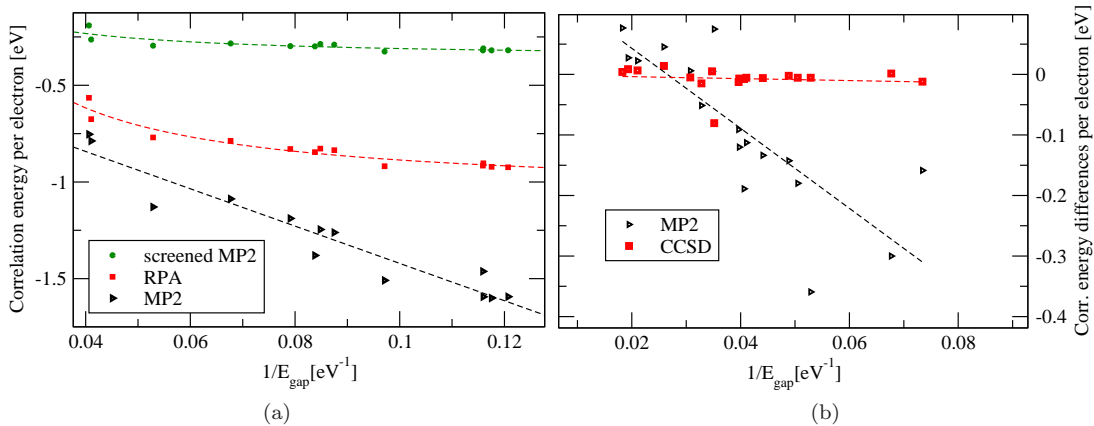

\subfloat[]{\includegraphics[width=0.4\textwidth,clip=true]{fig2a.eps}}
\hspace{0.01cm}
\subfloat[]{\includegraphics[width=0.4\textwidth,clip=true]{fig2b.eps}}
\caption{(Colour online) (a) Screened MP2, unscreened MP2 and RPA energies for a variety of finite-basis simulation-cell electron
gases with electron numbers $N=14-730$ corresponding to closed-shell configurations of a
simple cubic reciprocal-space lattice. (b) Differences between the RPA and MP2/CCSD correlation energies 
($r_s=1.0$~a.u., $\gamma=\sqrt{2}$) show that MP2 is divergent as the band gap
closes (on approach to the thermodynamic limit). CCSD is convergent to a constant energy
offset with respect to RPA which is only serendipitously close to zero for this $r_s$ and $\gamma$ \footnote{This agreement is the cross-over between CCSD and RPA due to overcorrelation on approach to the complete basis set limit similar to that seen in Fig. 5(b) in Ref.~\onlinecite{Shepherd2012c}.}.
}
\label{fig:1}
\end{figure*}

MP2 correlation energies per electron are presented in \reffig{fig:1} for sets of
finite-$N$ electron gases.
As the electron number increases the HF gap becomes smaller.
Concomitantly the MP2 correlation energy per electron increases linearly with the decreasing HF gap.
This conclusively demonstrates
that our approach recovers the expected divergence and physical behaviour from this method.
We stress that any approximate method suitable for metals and solids in general is required to
yield correlation energies per electron that converge to a constant in the $N\rightarrow\infty$ limit.
To further validate this as an approach accurately capturing the TDL, we compare this
with the finite-basis electron gas energies from identically constructed random-phase approximation (RPA) calculations,
which show a convergent behaviour (with a finite-size error as $\sim N^{-1}$~\cite{Drummond2008})
as anticipated. All RPA results in this work are calculated using a HF reference.
We note that the RPA referred to in this work corresponds to the so-called direct
RPA in which the employed two-electron integrals are not antisymmetrized.

We now return to the divergence of the MP2 correlation energy and show that it can be lifted
by replacing the bare Ewald interaction with a screened interaction in Eq.~(\ref{eq:mp2amplitudes}).
In this approach the ``screened MP2'' amplitudes are given by
\begin{equation}
t_{ij}^{ab}=\frac{\langle \Psi_{ij}^{ab} | \hat{v}_{\rm TF} | \Psi^{\rm HF} \rangle}{\epsilon_i+\epsilon_j-\epsilon_a-\epsilon_b},
\label{eq:scmp2amplitudes}
\end{equation}
where $\hat{v}_{\rm TF}$ refers to the Thomas--Fermi screened Coulomb interaction.
These amplitudes yield correlation energies per electron that converge for metallic systems with a rate similar to RPA, as shown in Fig.~\ref{fig:1}.
However, the screened MP2 energies strongly underestimate
the true correlation energy, as will become clear later.
We note that the introduction of the Thomas--Fermi screening is difficult to motivate in M\o ller-Plesset
perturbation theory. In Hedin's $GW$ theory, however, this corresponds to a static approximation
of the frequency-dependent screened electron interaction $W$ calculated in the RPA~\cite{Hedin1965}.
As such, our choice of screening has two advantages: (i) for homogeneous systems, the screening length depends on the electronic density only and the screened interaction is readily given by ${v}_{\rm TF}(r)=e^{-k_0 r}/r$, where $k_0$ is the Thomas--Fermi wavevector, and (ii) in the case of inhomogeneous systems, one can employ the $W$ calculated in the RPA of already existing $GW$ implementations.

Having now numerically demonstrated the well-accepted behaviour for the MP2 energy, we turn our
attention towards another approximate and widely-used quantum chemical method
for which the behaviour on approach to the TDL is
unverified --- coupled cluster singles and doubles theory (CCSD).
The CCSD doubles amplitudes are obtained by solving the so-called amplitude equations
given by
$\langle \Psi_{ij}^{ab} | e^{-T} H e^{T} | \Psi^{\rm HF} \rangle = 0 $,
where $T=T_2=(2!)^{-2} \sum_{ij} \sum_{ab} t_{ij}^{ab} \hat{c}^{\dagger}_a \hat{c}^{\dagger}_b \hat{c}_j \hat{c}_i$ for the HEG~\cite{Hirata2004,Bartlett2007}.
We note in passing that CCSD and coupled-cluster doubles theory (CCD) are equivalent for the
homogeneous electron gas due to the complete absence of symmetry-allowed single-excitations
in its many-body expansion.
$\hat{c}^{\dagger}_n$ ($\hat{c}_n$) are electron creation (annihilation) operators.
There has been surprisingly little literature concerning coupled cluster theory
for solids, in spite of the wealth of applications they have received in
the molecular quantum chemistry community.
Even though there has been some discussion of CCSD with \emph{approximate} amplitude equations,
but these more closely resemble the RPA equations\cite{Freeman1977,Bishop1978,Bishop1982,Hirata2012}.
As such, to the best of our knowledge, the question of whether CCSD diverges in the
TDL for metallic systems has not yet been conclusively addressed. 

Due to the relatively expensive scaling of such methods, simulations of an $N=730$ electron gas
with current fully-periodic codes\cite{VASP} are prohibitively expensive. However, we have found
that further reduction in finite-size effects can be achieved by taking the difference between
the CCSD energy with the RPA energy and in this difference the limiting behaviour is
more clear due to cancellations in the $N^{-1}$ term. We have also taken advantage of other simulation-cell lattices (face-centered cubic
and body-centered cubic) to provide more closed-shell configurations. Taking energy differences
in this way allows us to clearly demonstrate, in \reffig{fig:1}, that the CCSD energy converges at the
same rate as RPA\footnote{The outliers at $\sim$0.035 and $\sim$0.075 eV$^{-1}$ are due to these (bcc) systems having spuriously low and high band gaps respectively due to the lattice shape}.
Even though CCSD is exact through third-order perturbation theory, it performs a resummation
of infinitely many contributions to the correlation energy
(for instance all bubble diagrams as in RPA) of higher-order terms that
lift the divergence of order-by-order perturbation theory for metals. Furthermore, CCSD includes ladder diagrams that are understood to be important in the description of correlation at low densities\cite{Mattuck1967}.

\begin{figure}
\includegraphics[width=0.45\textwidth,clip=true]{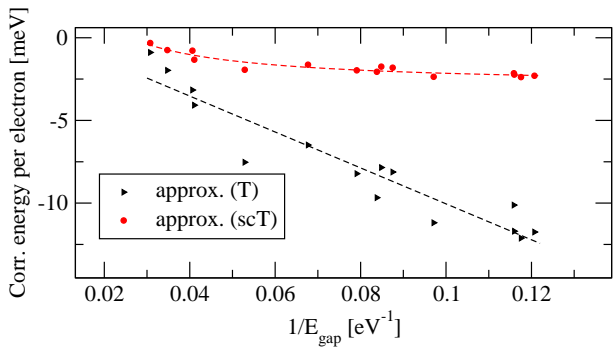}
\caption{(Colour online) Approximate screened and unscreened perturbative triples correlation energies per electron
for a range of finite-electron number calculations using $r_s$=1.0~a.u. and $\gamma=\sqrt{2}$.
}
\label{fig:4}
\end{figure}

Having demonstrated the convergent behaviour for the CCSD energy, we now turn our attention
to the perturbative triples (T) correction to CCSD~\cite{Raghavachari1989,Bartlett2007}.
CCSD(T) theory employs an exponential ansatz for the wavefunction given by $e^T \Psi_{\rm HF}$, where
$T=T_1+T_2+T_3$ and
$T_3=(3!)^{-2}\sum_{ijk} \sum_{abc} t_{ijk}^{abc} \hat{c}^{\dagger}_a \hat{c}^{\dagger}_b \hat{c}^{\dagger}_c \hat{c}_k \hat{c}_j \hat{c}_i$.
The corresponding triples amplitudes $t_{ijk}^{abc}$ are
calculated in a perturbative way reading
\begin{align}
t_{ijk}^{abc}
=&\frac{\langle \Psi_{ijk}^{abc} | [\hat{v},T_2] | \Psi^{\rm HF} \rangle }{\epsilon_i+\epsilon_j+\epsilon_k-\epsilon_a-\epsilon_b-\epsilon_c}.
\label{eq:t3}
\end{align}
Once obtained, the coupling of the triples with the doubles amplitudes is considered
only in an approximate fashion and its contribution to the CCSD correlation energy is calculated by
\begin{equation}
E^{\rm (T)}=\sum_{ij}\sum_{ab}t_{ij}^{ab}\langle \Psi_{ij}^{ab} | [H,T_3] | \Psi^{\rm HF} \rangle,
\label{eq:Ten}
\end{equation}
where the $t_{ij}^{ab}$ amplitudes are obtained from an underlying CCSD calculation.

Exploring the behaviour of Eq.~(\ref{eq:Ten}) for large electron numbers (N$>$200) is hindered
by the large computational cost.
As such we seek to investigate the behaviour of $E^{(T)}$ for metals by approximating Eq.~(\ref{eq:Ten})
in a manner that leaves the qualitative behaviour unchanged. We therefore approximate the CCSD amplitudes $t_{ij}^{ab}$ in the above expression by screened MP2
amplitudes from Eq.~(\ref{eq:scmp2amplitudes}). We have shown before that these amplitudes lead to a convergence
of the MP2 energy at the same rate as the CCSD and RPA energy.
Fig.~\ref{fig:4} shows that the approximate (T) energy expression from Eq.~(\ref{eq:Ten}), however, clearly diverges
for metals at the same rate as MP2.
From this we conclude that the full (T) contribution to the CCSD(T) correlation energy diverges as well
and that CCSD(T) is not a suitable method for metals.

We can now apply the same modification to the expression for the perturbative triples amplitudes in Eq.~(\ref{eq:t3})
as we have done for the screened MP2 amplitudes.
Replacing $\hat{v}$ in Eq.~(\ref{eq:t3}) with $\hat{v}_{\rm TF}$ yields
\begin{equation}
t_{ijk}^{abc}=\frac{\langle \Psi_{ijk}^{abc} | [\hat{v}_{\rm TF},T_2] | \Psi^{\rm HF} \rangle }{\epsilon_i+\epsilon_j+\epsilon_k-\epsilon_a-\epsilon_b-\epsilon_c}.
\end{equation}
Fig.~\ref{fig:4} demonstrates that the energies per electron calculated in the manner described above converges
for $N\rightarrow\infty$.
From this we conclude that the divergence for metallic systems of the full perturbative triples correction
can be lifted by using a screened interaction kernel for the evaluation of triples amplitudes.

\begin{figure}[t]
\includegraphics[width=0.4\textwidth,clip=true]{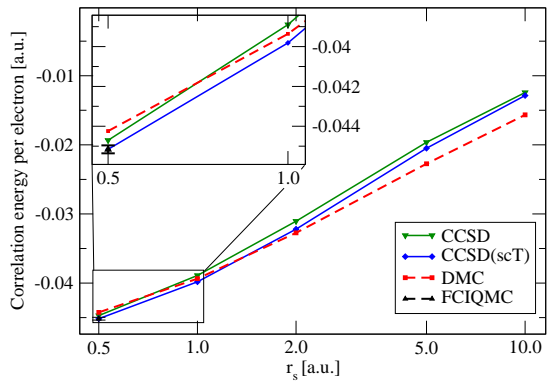}
\caption{(Colour online)
Complete basis set limit correlation energies per electron of DMC, CCSD, CCSD(scT) and FCIQMC for a range of densitites.
The exact FCIQMC and DMC SJ3 Backflow results are taken from Ref.~\cite{Shepherd2012a}
and Ref.~\cite{Rios2006}, respectively.
}
\label{fig:5}
\end{figure}

To test the accuracy of CCSD and CCSD(scT) (CCSD and screened perturbative triples) we calculate
the complete basis set limit correlation energies for the 54 electron system and compare to
diffusion Monte Carlo (DMC) and full configuration interaction quantum Monte Carlo (FCIQMC)
results for a range of realistic metallic densities.
The basis set extrapolations on the level of the wavefunction based methods
were carried out using M=700-1600 plane waves and the procedures outlined in Ref.~\cite{Shepherd2012c}.
Fig.~\ref{fig:5} shows the correlation energies of the different methods.
Our findings show that CCSD and CCSD(scT) become more accurate as the electronic density increases.
For very high densities ($r_{\rm s}=0.5$~a.u.) the quantum chemical wavefunction based methods
yield energies below the DMC ones. Although coupled-cluster methods are non-variational, the comparison to
exact FCIQMC results shows that CCSD and CCSD(scT) are closer to the exact energies in this density regime.
As the electronic density decreases the coupled-cluster methods capture less correlation energy.
We attribute this tendency to the increasing multi-reference character of the HEG wavefunction
at lower densities, which is difficult to treat with coupled cluster methods. This is likely
affected by the presence of a phase transition to the Wigner crystal, which however occurs at much lower densities than studied here~\cite{Drummond2004}.

\emph{Concluding remarks.} -- In summary, we have shown that a judicious choice of finite-size and finite-electron
number homogeneous electron gas models
can be used to demonstrate the limiting behaviour of the correlation energy in approximate many-body theories for
metallic systems with modest computational cost.
By comparing to RPA correlation energies we control for basis set incompleteness and finite-size errors.
As a first application of the outlined methodology, we have verified the divergence of MP2 energies in metals.
Furthermore we have shown that CCSD converges for metals at the same rate as RPA with respect to the system size $N$.
We have shown that the divergence in MP2 can be lifted by using a Thomas-Fermi screening in the calculation of MP2
amplitudes and yields a rate of convergence that is similar to the RPA and CCSD theory.
The screened MP2 approximation captures, however, only about half of the CCSD correlation energy in the same basis set.

The divergence of CCSD(T) for metals has been investigated by approximating the CCSD amplitudes with screened MP2 amplitudes.
We find that approximate (T) correlation energies per electron diverge as $N\rightarrow\infty$ and
conclude from this observation that the CCSD(T) method is not suitable for treating electron correlations in metals.
However, by the introduction of the Thomas-Fermi screening in the calculation of the triples amplitudes
this divergence can be lifted.
We demonstrate that the electronic correlation energies for the 54 electron system obtained using CCSD plus screened perturbative
triples (scT) are in good agreement with DMC results despite showing a tendency to capture less correlation for lower densities.
Going beyond the electron gas, it should be possible to model the screening with pre-existing \emph{ab initio} $GW$ codes
(e.g. Ref.~\onlinecite{VASP}).

In the high-density regime ($r_s \leq$ 1.5~a.u.) of the electron gas, CCSD(scT) recovers the correlation energy to high accuracy.
Even though it is well understood that in the high-density \emph{limit} the RPA plus a correction
from second-order exchange yields highly accurate results for the electron gas~\cite{Nozieres1958,Freeman1977},
the accuracy of these methods is not transferable to real systems~\cite{Grueneis2009,Paier2012}.
This is in contrast to
CC methods that are amongst the only wavefunction based methods that routinely achieve chemical accuracy with
polynomially-scaling computational cost~\cite{Bartlett2007}.
It is hoped that the developments presented here will be of interest to the
growing field aiming to study condensed matter systems with these techniques.

\emph{Acknowledgements.} -- The authors thank Ali Alavi, Georg Kresse, Martijn Marsman, Alex J. W. Thom and Tom Henderson for discussions. One of us (AG) gratefully acknowledges an APART-fellowship of the Austrian Academy of Sciences, and the other (JJS) EPSRC for funding.

\bibliography{divbib}

\begin{thebibliography}{38}%
\makeatletter
\providecommand \@ifxundefined [1]{%
 \@ifx{#1\undefined}
}%
\providecommand \@ifnum [1]{%
 \ifnum #1\expandafter \@firstoftwo
 \else \expandafter \@secondoftwo
 \fi
}%
\providecommand \@ifx [1]{%
 \ifx #1\expandafter \@firstoftwo
 \else \expandafter \@secondoftwo
 \fi
}%
\providecommand \natexlab [1]{#1}%
\providecommand \enquote  [1]{``#1''}%
\providecommand \bibnamefont  [1]{#1}%
\providecommand \bibfnamefont [1]{#1}%
\providecommand \citenamefont [1]{#1}%
\providecommand \href@noop [0]{\@secondoftwo}%
\providecommand \href [0]{\begingroup \@sanitize@url \@href}%
\providecommand \@href[1]{\@@startlink{#1}\@@href}%
\providecommand \@@href[1]{\endgroup#1\@@endlink}%
\providecommand \@sanitize@url [0]{\catcode `\\12\catcode `\$12\catcode
  `\&12\catcode `\#12\catcode `\^12\catcode `\_12\catcode `\%12\relax}%
\providecommand \@@startlink[1]{}%
\providecommand \@@endlink[0]{}%
\providecommand \url  [0]{\begingroup\@sanitize@url \@url }%
\providecommand \@url [1]{\endgroup\@href {#1}{\urlprefix }}%
\providecommand \urlprefix  [0]{URL }%
\providecommand \Eprint [0]{\href }%
\providecommand \doibase [0]{http://dx.doi.org/}%
\providecommand \selectlanguage [0]{\@gobble}%
\providecommand \bibinfo  [0]{\@secondoftwo}%
\providecommand \bibfield  [0]{\@secondoftwo}%
\providecommand \translation [1]{[#1]}%
\providecommand \BibitemOpen [0]{}%
\providecommand \bibitemStop [0]{}%
\providecommand \bibitemNoStop [0]{.\EOS\space}%
\providecommand \EOS [0]{\spacefactor3000\relax}%
\providecommand \BibitemShut  [1]{\csname bibitem#1\endcsname}%
\let\auto@bib@innerbib\@empty
\bibitem [{\citenamefont {Booth}\ \emph {et~al.}(2013)\citenamefont {Booth},
  \citenamefont {Gr\"{u}neis}, \citenamefont {Kresse},\ and\ \citenamefont
  {Alavi}}]{Booth2013}%
  \BibitemOpen
  \bibfield  {author} {\bibinfo {author} {\bibfnamefont {G.~H.}\ \bibnamefont
  {Booth}}, \bibinfo {author} {\bibfnamefont {A.}~\bibnamefont {Gr\"{u}neis}},
  \bibinfo {author} {\bibfnamefont {G.}~\bibnamefont {Kresse}}, \ and\ \bibinfo
  {author} {\bibfnamefont {A.}~\bibnamefont {Alavi}},\ }\href@noop {}
  {\bibfield  {journal} {\bibinfo  {journal} {Nature}\ }\textbf {\bibinfo
  {volume} {493}},\ \bibinfo {pages} {365} (\bibinfo {year}
  {2013})}\BibitemShut {NoStop}%
\bibitem [{\citenamefont {Del~Ben}\ \emph {et~al.}(2012)\citenamefont
  {Del~Ben}, \citenamefont {Hutter},\ and\ \citenamefont
  {VandeVondele}}]{Ben2012}%
  \BibitemOpen
  \bibfield  {author} {\bibinfo {author} {\bibfnamefont {M.}~\bibnamefont
  {Del~Ben}}, \bibinfo {author} {\bibfnamefont {J.}~\bibnamefont {Hutter}}, \
  and\ \bibinfo {author} {\bibfnamefont {J.}~\bibnamefont {VandeVondele}},\
  }\href@noop {} {\bibfield  {journal} {\bibinfo  {journal} {Journal of
  Chemical Theory and Computation}\ }\textbf {\bibinfo {volume} {8}},\ \bibinfo
  {pages} {4177} (\bibinfo {year} {2012})}\BibitemShut {NoStop}%
\bibitem [{\citenamefont {M\"{u}ller}\ and\ \citenamefont
  {Paulus}(2012)}]{Paulus2012}%
  \BibitemOpen
  \bibfield  {author} {\bibinfo {author} {\bibfnamefont {C.}~\bibnamefont
  {M\"{u}ller}}\ and\ \bibinfo {author} {\bibfnamefont {B.}~\bibnamefont
  {Paulus}},\ }\href@noop {} {\bibfield  {journal} {\bibinfo  {journal} {Phys.
  Chem. Chem. Phys.}\ }\textbf {\bibinfo {volume} {14}},\ \bibinfo {pages}
  {7605} (\bibinfo {year} {2012})}\BibitemShut {NoStop}%
\bibitem [{\citenamefont {Maschio}\ \emph {et~al.}(2007)\citenamefont
  {Maschio}, \citenamefont {Usvyat}, \citenamefont {Manby}, \citenamefont
  {Casassa}, \citenamefont {Pisani},\ and\ \citenamefont
  {Sch\"{u}tz}}]{Schutz2007}%
  \BibitemOpen
  \bibfield  {author} {\bibinfo {author} {\bibfnamefont {L.}~\bibnamefont
  {Maschio}}, \bibinfo {author} {\bibfnamefont {D.}~\bibnamefont {Usvyat}},
  \bibinfo {author} {\bibfnamefont {F.~R.}\ \bibnamefont {Manby}}, \bibinfo
  {author} {\bibfnamefont {S.}~\bibnamefont {Casassa}}, \bibinfo {author}
  {\bibfnamefont {C.}~\bibnamefont {Pisani}}, \ and\ \bibinfo {author}
  {\bibfnamefont {M.}~\bibnamefont {Sch\"{u}tz}},\ }\href@noop {} {\bibfield
  {journal} {\bibinfo  {journal} {Phys. Rev. B}\ ,\ \bibinfo {pages} {075101}}
  (\bibinfo {year} {2007})}\BibitemShut {NoStop}%
\bibitem [{\citenamefont {Casassa}\ \emph {et~al.}(2007)\citenamefont
  {Casassa}, \citenamefont {Halo}, \citenamefont {Maschio}, \citenamefont
  {Roetti},\ and\ \citenamefont {Pisani}}]{Pisani2007}%
  \BibitemOpen
  \bibfield  {author} {\bibinfo {author} {\bibfnamefont {S.}~\bibnamefont
  {Casassa}}, \bibinfo {author} {\bibfnamefont {M.}~\bibnamefont {Halo}},
  \bibinfo {author} {\bibfnamefont {L.}~\bibnamefont {Maschio}}, \bibinfo
  {author} {\bibfnamefont {C.}~\bibnamefont {Roetti}}, \ and\ \bibinfo {author}
  {\bibfnamefont {C.}~\bibnamefont {Pisani}},\ }\href@noop {} {\bibfield
  {journal} {\bibinfo  {journal} {Theor. Chem. Acc.}\ }\textbf {\bibinfo
  {volume} {117}},\ \bibinfo {pages} {781} (\bibinfo {year}
  {2007})}\BibitemShut {NoStop}%
\bibitem [{\citenamefont {Ayala}\ \emph {et~al.}(2001)\citenamefont {Ayala},
  \citenamefont {Kudin},\ and\ \citenamefont {Scuseria}}]{Scuseria2001}%
  \BibitemOpen
  \bibfield  {author} {\bibinfo {author} {\bibfnamefont {P.}~\bibnamefont
  {Ayala}}, \bibinfo {author} {\bibfnamefont {K.}~\bibnamefont {Kudin}}, \ and\
  \bibinfo {author} {\bibfnamefont {G.}~\bibnamefont {Scuseria}},\ }\href@noop
  {} {\bibfield  {journal} {\bibinfo  {journal} {J. Chem. Phys.}\ }\textbf
  {\bibinfo {volume} {115}},\ \bibinfo {pages} {9698} (\bibinfo {year}
  {2001})}\BibitemShut {NoStop}%
\bibitem [{\citenamefont {Usvyat}\ \emph {et~al.}(2011)\citenamefont {Usvyat},
  \citenamefont {Civalleri}, \citenamefont {Maschio}, \citenamefont {Dovesi},
  \citenamefont {Pisani},\ and\ \citenamefont {Sch\"{u}tz}}]{Schutz2011}%
  \BibitemOpen
  \bibfield  {author} {\bibinfo {author} {\bibfnamefont {D.}~\bibnamefont
  {Usvyat}}, \bibinfo {author} {\bibfnamefont {B.}~\bibnamefont {Civalleri}},
  \bibinfo {author} {\bibfnamefont {L.}~\bibnamefont {Maschio}}, \bibinfo
  {author} {\bibfnamefont {R.}~\bibnamefont {Dovesi}}, \bibinfo {author}
  {\bibfnamefont {C.}~\bibnamefont {Pisani}}, \ and\ \bibinfo {author}
  {\bibfnamefont {M.}~\bibnamefont {Sch\"{u}tz}},\ }\href@noop {} {\bibfield
  {journal} {\bibinfo  {journal} {J. Chem. Phys.}\ }\textbf {\bibinfo {volume}
  {134}},\ \bibinfo {eid} {214105} (\bibinfo {year} {2011})}\BibitemShut
  {NoStop}%
\bibitem [{\citenamefont {Marsman}\ \emph {et~al.}(2009)\citenamefont
  {Marsman}, \citenamefont {Gr\"{u}neis}, \citenamefont {Paier},\ and\
  \citenamefont {Kresse}}]{Kresse2009}%
  \BibitemOpen
  \bibfield  {author} {\bibinfo {author} {\bibfnamefont {M.}~\bibnamefont
  {Marsman}}, \bibinfo {author} {\bibfnamefont {A.}~\bibnamefont
  {Gr\"{u}neis}}, \bibinfo {author} {\bibfnamefont {J.}~\bibnamefont {Paier}},
  \ and\ \bibinfo {author} {\bibfnamefont {G.}~\bibnamefont {Kresse}},\ }\href
  {\doibase 10.1063/1.3126249} {\bibfield  {journal} {\bibinfo  {journal} {J.
  Chem. Phys.}\ }\textbf {\bibinfo {volume} {130}},\ \bibinfo {eid} {184103}
  (\bibinfo {year} {2009})}\BibitemShut {NoStop}%
\bibitem [{\citenamefont {Stoll}\ and\ \citenamefont {Doll}(2012)}]{Stoll2012}%
  \BibitemOpen
  \bibfield  {author} {\bibinfo {author} {\bibfnamefont {H.}~\bibnamefont
  {Stoll}}\ and\ \bibinfo {author} {\bibfnamefont {K.}~\bibnamefont {Doll}},\
  }\href@noop {} {\bibfield  {journal} {\bibinfo  {journal} {J. Chem. Phys.}\
  ,\ \bibinfo {pages} {074106}} (\bibinfo {year} {2012})}\BibitemShut {NoStop}%
\bibitem [{\citenamefont {Nolan}\ \emph {et~al.}(2009)\citenamefont {Nolan},
  \citenamefont {Gillan}, \citenamefont {Alf\`e}, \citenamefont {Allan},\ and\
  \citenamefont {Manby}}]{Manby2009}%
  \BibitemOpen
  \bibfield  {author} {\bibinfo {author} {\bibfnamefont {S.~J.}\ \bibnamefont
  {Nolan}}, \bibinfo {author} {\bibfnamefont {M.~J.}\ \bibnamefont {Gillan}},
  \bibinfo {author} {\bibfnamefont {D.}~\bibnamefont {Alf\`e}}, \bibinfo
  {author} {\bibfnamefont {N.~L.}\ \bibnamefont {Allan}}, \ and\ \bibinfo
  {author} {\bibfnamefont {F.~R.}\ \bibnamefont {Manby}},\ }\href@noop {}
  {\bibfield  {journal} {\bibinfo  {journal} {Phys. Rev. B}\ }\textbf {\bibinfo
  {volume} {80}},\ \bibinfo {pages} {165109} (\bibinfo {year}
  {2009})}\BibitemShut {NoStop}%
\bibitem [{\citenamefont {Binnie}\ \emph {et~al.}(2010)\citenamefont {Binnie},
  \citenamefont {Nolan}, \citenamefont {Drummond}, \citenamefont {Alf\`e},
  \citenamefont {Allan}, \citenamefont {Manby},\ and\ \citenamefont
  {Gillan}}]{Gillan2010}%
  \BibitemOpen
  \bibfield  {author} {\bibinfo {author} {\bibfnamefont {S.~J.}\ \bibnamefont
  {Binnie}}, \bibinfo {author} {\bibfnamefont {S.~J.}\ \bibnamefont {Nolan}},
  \bibinfo {author} {\bibfnamefont {N.~D.}\ \bibnamefont {Drummond}}, \bibinfo
  {author} {\bibfnamefont {D.}~\bibnamefont {Alf\`e}}, \bibinfo {author}
  {\bibfnamefont {N.~L.}\ \bibnamefont {Allan}}, \bibinfo {author}
  {\bibfnamefont {F.~R.}\ \bibnamefont {Manby}}, \ and\ \bibinfo {author}
  {\bibfnamefont {M.~J.}\ \bibnamefont {Gillan}},\ }\href@noop {} {\bibfield
  {journal} {\bibinfo  {journal} {Phys. Rev. B}\ }\textbf {\bibinfo {volume}
  {82}} (\bibinfo {year} {2010})}\BibitemShut {NoStop}%
\bibitem [{\citenamefont {P.Schwerdtfeger}\ \emph {et~al.}(2010)\citenamefont
  {P.Schwerdtfeger}, \citenamefont {B.Assadollahzadeh},\ and\ \citenamefont
  {A.Hermann}}]{Schwerdtfeger2010}%
  \BibitemOpen
  \bibfield  {author} {\bibinfo {author} {\bibnamefont {P.Schwerdtfeger}},
  \bibinfo {author} {\bibnamefont {B.Assadollahzadeh}}, \ and\ \bibinfo
  {author} {\bibnamefont {A.Hermann}},\ }\href@noop {} {\bibfield  {journal}
  {\bibinfo  {journal} {Phys. Rev. B}\ }\textbf {\bibinfo {volume} {82}},\
  \bibinfo {pages} {205111} (\bibinfo {year} {2010})}\BibitemShut {NoStop}%
\bibitem [{\citenamefont {Marsman}\ \emph {et~al.}(2008)\citenamefont
  {Marsman}, \citenamefont {Paier}, \citenamefont {Stroppa},\ and\
  \citenamefont {Kresse}}]{Marsman2008}%
  \BibitemOpen
  \bibfield  {author} {\bibinfo {author} {\bibfnamefont {M.}~\bibnamefont
  {Marsman}}, \bibinfo {author} {\bibfnamefont {J.}~\bibnamefont {Paier}},
  \bibinfo {author} {\bibfnamefont {A.}~\bibnamefont {Stroppa}}, \ and\
  \bibinfo {author} {\bibfnamefont {G.}~\bibnamefont {Kresse}},\ }\href@noop {}
  {\bibfield  {journal} {\bibinfo  {journal} {J. Phys.: Condens. Matter}\
  }\textbf {\bibinfo {volume} {20}},\ \bibinfo {pages} {064201} (\bibinfo
  {year} {2008})}\BibitemShut {NoStop}%
\bibitem [{\citenamefont {Shepherd}\ \emph
  {et~al.}(2012{\natexlab{a}})\citenamefont {Shepherd}, \citenamefont {Booth},
  \citenamefont {Gr\"{u}neis},\ and\ \citenamefont {Alavi}}]{Shepherd2012a}%
  \BibitemOpen
  \bibfield  {author} {\bibinfo {author} {\bibfnamefont {J.~J.}\ \bibnamefont
  {Shepherd}}, \bibinfo {author} {\bibfnamefont {G.}~\bibnamefont {Booth}},
  \bibinfo {author} {\bibfnamefont {A.}~\bibnamefont {Gr\"{u}neis}}, \ and\
  \bibinfo {author} {\bibfnamefont {A.}~\bibnamefont {Alavi}},\ }\href@noop {}
  {\bibfield  {journal} {\bibinfo  {journal} {Phys. Rev. B}\ }\textbf {\bibinfo
  {volume} {85}},\ \bibinfo {pages} {081103} (\bibinfo {year}
  {2012}{\natexlab{a}})}\BibitemShut {NoStop}%
\bibitem [{\citenamefont {Shepherd}\ \emph
  {et~al.}(2012{\natexlab{b}})\citenamefont {Shepherd}, \citenamefont {Booth},\
  and\ \citenamefont {Alavi}}]{Shepherd2012b}%
  \BibitemOpen
  \bibfield  {author} {\bibinfo {author} {\bibfnamefont {J.~J.}\ \bibnamefont
  {Shepherd}}, \bibinfo {author} {\bibfnamefont {G.~H.}\ \bibnamefont {Booth}},
  \ and\ \bibinfo {author} {\bibfnamefont {A.}~\bibnamefont {Alavi}},\
  }\href@noop {} {\bibfield  {journal} {\bibinfo  {journal} {J. Chem. Phys.}\
  }\textbf {\bibinfo {volume} {136}},\ \bibinfo {pages} {244101} (\bibinfo
  {year} {2012}{\natexlab{b}})}\BibitemShut {NoStop}%
\bibitem [{\citenamefont {Shepherd}\ \emph
  {et~al.}(2012{\natexlab{c}})\citenamefont {Shepherd}, \citenamefont
  {Gr\"{u}neis}, \citenamefont {Booth}, \citenamefont {Kresse},\ and\
  \citenamefont {Alavi}}]{Shepherd2012c}%
  \BibitemOpen
  \bibfield  {author} {\bibinfo {author} {\bibfnamefont {J.~J.}\ \bibnamefont
  {Shepherd}}, \bibinfo {author} {\bibfnamefont {A.}~\bibnamefont
  {Gr\"{u}neis}}, \bibinfo {author} {\bibfnamefont {G.~H.}\ \bibnamefont
  {Booth}}, \bibinfo {author} {\bibfnamefont {G.}~\bibnamefont {Kresse}}, \
  and\ \bibinfo {author} {\bibfnamefont {A.}~\bibnamefont {Alavi}},\
  }\href@noop {} {\bibfield  {journal} {\bibinfo  {journal} {Phys. Rev. B.}\
  }\textbf {\bibinfo {volume} {86}},\ \bibinfo {pages} {035111} (\bibinfo
  {year} {2012}{\natexlab{c}})}\BibitemShut {NoStop}%
\bibitem [{\citenamefont {Ewald}(1921)}]{Ewald1921}%
  \BibitemOpen
  \bibfield  {author} {\bibinfo {author} {\bibfnamefont {P.}~\bibnamefont
  {Ewald}},\ }\href@noop {} {\bibfield  {journal} {\bibinfo  {journal} {Ann.
  Phys.}\ }\textbf {\bibinfo {volume} {64}},\ \bibinfo {pages} {253} (\bibinfo
  {year} {1921})}\BibitemShut {NoStop}%
\bibitem [{\citenamefont {Fraser}\ \emph {et~al.}(1996)\citenamefont {Fraser},
  \citenamefont {Foulkes}, \citenamefont {Rajagopal}, \citenamefont {Needs},
  \citenamefont {Kenny},\ and\ \citenamefont {Williamson}}]{Fraser1996}%
  \BibitemOpen
  \bibfield  {author} {\bibinfo {author} {\bibfnamefont {L.~M.}\ \bibnamefont
  {Fraser}}, \bibinfo {author} {\bibfnamefont {W.~M.~C.}\ \bibnamefont
  {Foulkes}}, \bibinfo {author} {\bibfnamefont {G.}~\bibnamefont {Rajagopal}},
  \bibinfo {author} {\bibfnamefont {R.~J.}\ \bibnamefont {Needs}}, \bibinfo
  {author} {\bibfnamefont {S.~D.}\ \bibnamefont {Kenny}}, \ and\ \bibinfo
  {author} {\bibfnamefont {A.~J.}\ \bibnamefont {Williamson}},\ }\href@noop {}
  {\bibfield  {journal} {\bibinfo  {journal} {Phys. Rev. B}\ }\textbf {\bibinfo
  {volume} {53}},\ \bibinfo {pages} {1814} (\bibinfo {year}
  {1996})}\BibitemShut {NoStop}%
\bibitem [{\citenamefont {Martin}(2004)}]{Martin2004}%
  \BibitemOpen
  \bibfield  {author} {\bibinfo {author} {\bibfnamefont {R.}~\bibnamefont
  {Martin}},\ }\href@noop {} {\emph {\bibinfo {title} {Electronic Structure}}}\
  (\bibinfo  {publisher} {Cambridge University Press},\ \bibinfo {year}
  {2004})\BibitemShut {NoStop}%
\bibitem [{\citenamefont {Drummond}\ \emph {et~al.}(2008)\citenamefont
  {Drummond}, \citenamefont {Needs}, \citenamefont {Sorouri},\ and\
  \citenamefont {Foulkes}}]{Drummond2008}%
  \BibitemOpen
  \bibfield  {author} {\bibinfo {author} {\bibfnamefont {N.}~\bibnamefont
  {Drummond}}, \bibinfo {author} {\bibfnamefont {R.}~\bibnamefont {Needs}},
  \bibinfo {author} {\bibfnamefont {A.}~\bibnamefont {Sorouri}}, \ and\
  \bibinfo {author} {\bibfnamefont {W.}~\bibnamefont {Foulkes}},\ }\href@noop
  {} {\bibfield  {journal} {\bibinfo  {journal} {Phys. Rev. B}\ }\textbf
  {\bibinfo {volume} {78}},\ \bibinfo {pages} {125106} (\bibinfo {year}
  {2008})}\BibitemShut {NoStop}%
\bibitem [{Note1()}]{Note1}%
  \BibitemOpen
  \bibinfo {note} {Taking a \protect \emph {finite-basis} approach to the
  homogeneous electron gas has only recently been introduced to the
  literature~\cite {Shepherd2012a,Shepherd2012b,Shepherd2012c}}\BibitemShut
  {NoStop}%
\bibitem [{\citenamefont {Gr\"{u}neis}\ \emph {et~al.}(2010)\citenamefont
  {Gr\"{u}neis}, \citenamefont {Marsman},\ and\ \citenamefont
  {Kresse}}]{Gruneis2010}%
  \BibitemOpen
  \bibfield  {author} {\bibinfo {author} {\bibfnamefont {A.}~\bibnamefont
  {Gr\"{u}neis}}, \bibinfo {author} {\bibfnamefont {M.}~\bibnamefont
  {Marsman}}, \ and\ \bibinfo {author} {\bibfnamefont {G.}~\bibnamefont
  {Kresse}},\ }\href@noop {} {\bibfield  {journal} {\bibinfo  {journal} {J.
  Chem. Phys.}\ }\textbf {\bibinfo {volume} {133}},\ \bibinfo {eid} {074107}
  (\bibinfo {year} {2010})}\BibitemShut {NoStop}%
\bibitem [{\citenamefont {Hedin}(1965)}]{Hedin1965}%
  \BibitemOpen
  \bibfield  {author} {\bibinfo {author} {\bibfnamefont {L.}~\bibnamefont
  {Hedin}},\ }\href@noop {} {\bibfield  {journal} {\bibinfo  {journal} {Phys.
  Rev.}\ }\textbf {\bibinfo {volume} {139}},\ \bibinfo {pages} {A796} (\bibinfo
  {year} {1965})}\BibitemShut {NoStop}%
\bibitem [{\citenamefont {S.Hirata}\ \emph {et~al.}(2004)\citenamefont
  {S.Hirata}, \citenamefont {R.Podeszwa}, \citenamefont {M.Tobita},\ and\
  \citenamefont {R.J.Bartlett}}]{Hirata2004}%
  \BibitemOpen
  \bibfield  {author} {\bibinfo {author} {\bibnamefont {S.Hirata}}, \bibinfo
  {author} {\bibnamefont {R.Podeszwa}}, \bibinfo {author} {\bibnamefont
  {M.Tobita}}, \ and\ \bibinfo {author} {\bibnamefont {R.J.Bartlett}},\
  }\href@noop {} {\bibfield  {journal} {\bibinfo  {journal} {J. Chem. Phys.}\
  }\textbf {\bibinfo {volume} {120}},\ \bibinfo {pages} {2581} (\bibinfo {year}
  {2004})}\BibitemShut {NoStop}%
\bibitem [{\citenamefont {Bartlett}\ and\ \citenamefont
  {Musial}(2007)}]{Bartlett2007}%
  \BibitemOpen
  \bibfield  {author} {\bibinfo {author} {\bibfnamefont {R.~J.}\ \bibnamefont
  {Bartlett}}\ and\ \bibinfo {author} {\bibfnamefont {M.}~\bibnamefont
  {Musial}},\ }\href@noop {} {\bibfield  {journal} {\bibinfo  {journal} {Rev.
  Mod. Phys.}\ }\textbf {\bibinfo {volume} {79}},\ \bibinfo {pages} {291}
  (\bibinfo {year} {2007})}\BibitemShut {NoStop}%
\bibitem [{\citenamefont {Freeman}(1977)}]{Freeman1977}%
  \BibitemOpen
  \bibfield  {author} {\bibinfo {author} {\bibfnamefont {D.}~\bibnamefont
  {Freeman}},\ }\href@noop {} {\bibfield  {journal} {\bibinfo  {journal} {Phys.
  Rev. B}\ }\textbf {\bibinfo {volume} {15}},\ \bibinfo {pages} {5512}
  (\bibinfo {year} {1977})}\BibitemShut {NoStop}%
\bibitem [{\citenamefont {Bishop}\ and\ \citenamefont
  {L\"uhrmann}(1978)}]{Bishop1978}%
  \BibitemOpen
  \bibfield  {author} {\bibinfo {author} {\bibfnamefont {R.~F.}\ \bibnamefont
  {Bishop}}\ and\ \bibinfo {author} {\bibfnamefont {K.~H.}\ \bibnamefont
  {L\"uhrmann}},\ }\href@noop {} {\bibfield  {journal} {\bibinfo  {journal}
  {Phys. Rev. B.}\ }\textbf {\bibinfo {volume} {17}},\ \bibinfo {pages} {3757}
  (\bibinfo {year} {1978})}\BibitemShut {NoStop}%
\bibitem [{\citenamefont {Bishop}\ and\ \citenamefont
  {L\"uhrmann}(1982)}]{Bishop1982}%
  \BibitemOpen
  \bibfield  {author} {\bibinfo {author} {\bibfnamefont {R.~F.}\ \bibnamefont
  {Bishop}}\ and\ \bibinfo {author} {\bibfnamefont {K.~H.}\ \bibnamefont
  {L\"uhrmann}},\ }\href@noop {} {\bibfield  {journal} {\bibinfo  {journal}
  {Phys. Rev. B.}\ }\textbf {\bibinfo {volume} {26}},\ \bibinfo {pages} {5523}
  (\bibinfo {year} {1982})}\BibitemShut {NoStop}%
\bibitem [{\citenamefont {Hirata}\ and\ \citenamefont
  {Ohnishi}(2012)}]{Hirata2012}%
  \BibitemOpen
  \bibfield  {author} {\bibinfo {author} {\bibfnamefont {S.}~\bibnamefont
  {Hirata}}\ and\ \bibinfo {author} {\bibfnamefont {Y.}~\bibnamefont
  {Ohnishi}},\ }\href@noop {} {\bibfield  {journal} {\bibinfo  {journal} {Phys.
  Chem. Chem. Phys.}\ }\textbf {\bibinfo {volume} {14}},\ \bibinfo {pages}
  {7800} (\bibinfo {year} {2012})}\BibitemShut {NoStop}%
\bibitem [{VAS(2013)}]{VASP}%
  \BibitemOpen
  \href {http://www.vasp.at} {}\bibinfo {howpublished} {http://www.vasp.at}
  (\bibinfo {year} {2013})\BibitemShut {NoStop}%
\bibitem [{Note2()}]{Note2}%
  \BibitemOpen
  \bibinfo {note} {The outliers at $\sim $0.035 and $\sim $0.075 eV$^{-1}$ are
  due to these (bcc) systems having spuriously low and high band gaps
  respectively due to the lattice shape}\BibitemShut {NoStop}%
\bibitem [{\citenamefont {Mattuck}(1967)}]{Mattuck1967}%
  \BibitemOpen
  \bibfield  {author} {\bibinfo {author} {\bibfnamefont {R.~D.}\ \bibnamefont
  {Mattuck}},\ }\href@noop {} {\emph {\bibinfo {title} {A Guide to Feynman
  Diagrams in the Many-Body Problem}}}\ (\bibinfo  {publisher} {Dover},\
  \bibinfo {year} {1967})\BibitemShut {NoStop}%
\bibitem [{\citenamefont {K.Raghavachari}\ \emph {et~al.}(1989)\citenamefont
  {K.Raghavachari}, \citenamefont {G.W.Trucks}, \citenamefont {J.A.Pople},\
  and\ \citenamefont {M.Head-Gordon}}]{Raghavachari1989}%
  \BibitemOpen
  \bibfield  {author} {\bibinfo {author} {\bibnamefont {K.Raghavachari}},
  \bibinfo {author} {\bibnamefont {G.W.Trucks}}, \bibinfo {author}
  {\bibnamefont {J.A.Pople}}, \ and\ \bibinfo {author} {\bibnamefont
  {M.Head-Gordon}},\ }\href@noop {} {\bibfield  {journal} {\bibinfo  {journal}
  {Chem. Phys. Lett.}\ }\textbf {\bibinfo {volume} {157}},\ \bibinfo {pages}
  {479} (\bibinfo {year} {1989})}\BibitemShut {NoStop}%
\bibitem [{\citenamefont {Drummond}\ \emph {et~al.}(2004)\citenamefont
  {Drummond}, \citenamefont {Radnai}, \citenamefont {Trail}, \citenamefont
  {Towler},\ and\ \citenamefont {Needs}}]{Drummond2004}%
  \BibitemOpen
  \bibfield  {author} {\bibinfo {author} {\bibfnamefont {N.~D.}\ \bibnamefont
  {Drummond}}, \bibinfo {author} {\bibfnamefont {Z.}~\bibnamefont {Radnai}},
  \bibinfo {author} {\bibfnamefont {J.~R.}\ \bibnamefont {Trail}}, \bibinfo
  {author} {\bibfnamefont {M.~D.}\ \bibnamefont {Towler}}, \ and\ \bibinfo
  {author} {\bibfnamefont {R.~J.}\ \bibnamefont {Needs}},\ }\href@noop {}
  {\bibfield  {journal} {\bibinfo  {journal} {Phys. Rev. B}\ }\textbf {\bibinfo
  {volume} {69}},\ \bibinfo {pages} {085116} (\bibinfo {year}
  {2004})}\BibitemShut {NoStop}%
\bibitem [{\citenamefont {R\'{i}os}\ \emph {et~al.}(2006)\citenamefont
  {R\'{i}os}, \citenamefont {Ma}, \citenamefont {Drummond}, \citenamefont
  {Towler},\ and\ \citenamefont {Needs}}]{Rios2006}%
  \BibitemOpen
  \bibfield  {author} {\bibinfo {author} {\bibfnamefont {P.~L.}\ \bibnamefont
  {R\'{i}os}}, \bibinfo {author} {\bibfnamefont {A.}~\bibnamefont {Ma}},
  \bibinfo {author} {\bibfnamefont {N.~D.}\ \bibnamefont {Drummond}}, \bibinfo
  {author} {\bibfnamefont {M.~D.}\ \bibnamefont {Towler}}, \ and\ \bibinfo
  {author} {\bibfnamefont {R.~J.}\ \bibnamefont {Needs}},\ }\href@noop {}
  {\bibfield  {journal} {\bibinfo  {journal} {Phys. Rev. E}\ }\textbf {\bibinfo
  {volume} {74}},\ \bibinfo {pages} {066701} (\bibinfo {year}
  {2006})}\BibitemShut {NoStop}%
\bibitem [{\citenamefont {Nozi\`eres}\ and\ \citenamefont
  {Pines}(1958)}]{Nozieres1958}%
  \BibitemOpen
  \bibfield  {author} {\bibinfo {author} {\bibfnamefont {P.}~\bibnamefont
  {Nozi\`eres}}\ and\ \bibinfo {author} {\bibfnamefont {D.}~\bibnamefont
  {Pines}},\ }\href@noop {} {\bibfield  {journal} {\bibinfo  {journal} {Phys.
  Rev.}\ }\textbf {\bibinfo {volume} {111}},\ \bibinfo {pages} {442} (\bibinfo
  {year} {1958})}\BibitemShut {NoStop}%
\bibitem [{\citenamefont {Gr\"{u}neis}\ \emph {et~al.}(2009)\citenamefont
  {Gr\"{u}neis}, \citenamefont {Marsman}, \citenamefont {Harl}, \citenamefont
  {Schimka},\ and\ \citenamefont {Kresse}}]{Grueneis2009}%
  \BibitemOpen
  \bibfield  {author} {\bibinfo {author} {\bibfnamefont {A.}~\bibnamefont
  {Gr\"{u}neis}}, \bibinfo {author} {\bibfnamefont {M.}~\bibnamefont
  {Marsman}}, \bibinfo {author} {\bibfnamefont {J.}~\bibnamefont {Harl}},
  \bibinfo {author} {\bibfnamefont {L.}~\bibnamefont {Schimka}}, \ and\
  \bibinfo {author} {\bibfnamefont {G.}~\bibnamefont {Kresse}},\ }\href@noop {}
  {\bibfield  {journal} {\bibinfo  {journal} {J. Chem. Phys.}\ }\textbf
  {\bibinfo {volume} {131}},\ \bibinfo {pages} {154115} (\bibinfo {year}
  {2009})}\BibitemShut {NoStop}%
\bibitem [{\citenamefont {Paier}\ \emph {et~al.}(2012)\citenamefont {Paier},
  \citenamefont {Ren}, \citenamefont {Rinke}, \citenamefont {Scuseria},
  \citenamefont {Gr\"{u}neis}, \citenamefont {Kresse},\ and\ \citenamefont
  {Scheffler}}]{Paier2012}%
  \BibitemOpen
  \bibfield  {author} {\bibinfo {author} {\bibfnamefont {J.}~\bibnamefont
  {Paier}}, \bibinfo {author} {\bibfnamefont {X.}~\bibnamefont {Ren}}, \bibinfo
  {author} {\bibfnamefont {P.}~\bibnamefont {Rinke}}, \bibinfo {author}
  {\bibfnamefont {G.~E.}\ \bibnamefont {Scuseria}}, \bibinfo {author}
  {\bibfnamefont {A.}~\bibnamefont {Gr\"{u}neis}}, \bibinfo {author}
  {\bibfnamefont {G.}~\bibnamefont {Kresse}}, \ and\ \bibinfo {author}
  {\bibfnamefont {M.}~\bibnamefont {Scheffler}},\ }\href@noop {} {\bibfield
  {journal} {\bibinfo  {journal} {New Journal of Physics}\ }\textbf {\bibinfo
  {volume} {14}},\ \bibinfo {pages} {043002} (\bibinfo {year}
  {2012})}\BibitemShut {NoStop}%
\end{thebibliography}%

\end{document}